\documentclass[%
  reprint,
superscriptaddress,
nofootinbib,
 amsmath,amssymb,
 aps,
 prc,
 showkeys,
]{revtex4-1}

\usepackage{graphicx}% Include figure files
\usepackage{dcolumn}% Align table columns on decimal point
\usepackage{bm}% bold math
\usepackage{subfigure}
\usepackage{float} %the position of picture
\usepackage{hyperref}% add hypertext capabilities
\usepackage[mathlines]{lineno}% Enable numbering of text and display math
\usepackage{color}

\newcommand{\be}{\begin{equation}}
\newcommand{\ee}{\end{equation}}
\newcommand{\bea}{\begin{eqnarray}}
\newcommand{\eea}{\end{eqnarray}}

\begin{document}

\preprint{BeAGLE-BNL}

\title{Opportunities for Imaging Light Nuclei with a Second Interaction Region \\ at the Electron-Ion Collider}

\author{Wan~Chang}
\email{changwan@nynu.edu.cn}
\affiliation{%
 School of Physics and Electronic Engineering, Nanyang Normal University, Nanyang 473061, China
}%
\affiliation{%
 Key Laboratory of Quark and Lepton Physics (MOE) and Institute of Particle Physics, Central China Normal University, Wuhan 430079, China
}%

\author{Elke-Caroline~Aschenauer }
%\email{elke@bnl.gov}
\affiliation{%
 Department of Physics, Brookhaven National Laboratory, Upton, New York 11973, U.S.A.
}%

\author{Alexander~Jentsch}
\affiliation{%
 Department of Physics, Brookhaven National Laboratory, Upton, New York 11973, U.S.A.
}%

\author{Arjun~Kumar}
 \affiliation{Center for Nuclear Frontiers in Nuclear Science, Department of Physics and Astronomy, Stony Brook University, New York 11794-3800, USA}

\author{Zhoudunming~Tu}
\email{zhoudunming@bnl.gov}
\affiliation{%
 Department of Physics, Brookhaven National Laboratory, Upton, New York 11973, U.S.A.
}%
\affiliation{Center for Nuclear Frontiers in Nuclear Science, Department of Physics and Astronomy, Stony Brook University, New York 11794-3800, USA}

\author{Zhongbao~Yin}
\affiliation{%
 Key Laboratory of Quark and Lepton Physics (MOE) and Institute of Particle Physics, Central China Normal University, Wuhan 430079, China
}%

\date{\today}

\begin{abstract}
The upcoming Electron-Ion Collider (EIC) will address several outstanding puzzles in modern nuclear physics. Key questions—such as the partonic structure of nucleons and nuclei and the origin of their mass and spin—can be explored through high-energy electron–proton and electron–nucleus collisions. To maximize its scientific reach, the EIC community has advocated for the addition of a second interaction region equipped with a detector complementary to the EIC general purpose collider detector, ePIC. The pre-conceptual design of this interaction region aims to provide a different configuration from the first interaction region, which enhances forward acceptance at very small scattering angles ($\theta \sim 0$ mrad). This machine configuration would significantly benefit exclusive, tagging, and diffractive physics programs, complementing those of the ePIC experiment. In particular, accessing coherent diffractive processes on light nuclei by tagging of the full, intact nucleus is essential for mapping their spatial parton distributions. In this work, we present an exploratory study of the detection capabilities for light nuclei at a second EIC interaction region, with a detailed discussion of the accessible kinematic phase space and its implications for imaging.

\end{abstract}

\keywords{ EIC, IR-8, secondary focus, coherent diffractive events, detection efficiency}

\maketitle

\section{\label{sec:intro}Introduction}
The upcoming U.S.-based Electron Ion Collider (EIC) is being designed to collide electrons with various ions, ranging from protons to uranium, across a wide range of center-of-mass energies from 30 to 140 GeV, while achieving extremely high luminosity of $10^{33-34} \rm{cm}^{-2} \rm{sec}^{-1}$~\cite{ABDULKHALEK2022122447, ref:EICCDR, Accardi:2012qut}. Moreover, the EIC will facilitate collisions involving polarized electrons, protons, and light ions ($^{3}\rm{He}$)~\cite{ref:EICCDR}, making it the only collider in the world with this capability.

The EIC science program is extremely broad and diverse. It runs the gamut from detailed investigation of hadronic structure with unprecedented precision to explorations of new regimes of strongly interacting matter~\cite{Gelis:2010nm, Jalilian-Marian:2014ica}, providing an excellent way to explore the fundamental theory of Quantum Chromodynamics (QCD)\cite{Gaillard:1998ui, Oerter:2006iy}. The EIC will enable us to study the entire three-dimensional of partons inside the proton in a manner that surpasses what can be learned from standard collinear parton distributions, which only provide information about the longitudinal momentum structure. With the unique capability to span a wide range of momentum transfer $Q^{2}$ and Bjorken-$x$, the EIC can offer the most powerful tool to precisely quantify how the spin of gluons and that of quarks of various flavors contribute to the proton spin. Another area of EIC research that will be explored is nuclear partonic structure and how it might differ from a free proton. Finally, by utilizing the EIC's capability of colliding electrons with a broad spectrum of nuclear species, the EIC can serve as an irreplaceable frontier QCD laboratory for the systematic investigation of novel nuclear phenomena~\cite{Accardi:2012qut}. The production of vector mesons (VM) in coherent diffractive processes, $e+\rm{A} \rightarrow \it{e'}+\rm{VM}+\rm{A}'$, where $\rm{VM}$ =  $J/\psi$, $\phi$, $\rho$, is a unique process which has a primary focus at the EIC. Since only one final state particle is generated, the process is experimentally clean and can be unambiguously identified by the presence of a rapidity gap~\cite{Toll:2012mb, Abramowicz:1998ii}. What makes the coherent diffractive processes so interesting is that they are most sensitive to the underlying gluon distributions in nuclei~\cite{Jones:2013pga,  Jones:2016ldq, PhysRevD.106.074019}.

The EIC is designed to support two Interaction Regions at the ``6 o-clock" and ``8 o-clock" halls of the current Relativistic Heavy-Ion Collider (RHIC) facility (IR-6 and IR-8), although the EIC project scope only includes funding for a fully instrumented IR-6 and the associated ePIC~\cite{epic} detector. The presence of two general-purpose collider detectors supports the comprehensive EIC science program, facilitating cross-verification of significant findings and enhancing discoveries through complementarity. Furthermore, it enables the combination of different datasets from the two studies, a validated method for minimizing experimental systematic uncertainty.

In this paper, we use a pre-conceptual design of the EIC second interaction region, IR-8, and explore its capability to identify coherent electron-nucleus ($e$A) scattering events by tagging the full, intact nucleus. This study makes use of the major design characteristic of interest here, namely, a secondary focus. Simulations include the necessary far-forward (FF) detectors that are adapted to the proposed IR-8 hadron beam line geometry and its current magnetic field configuration. To measure the coherent heavy ions, one can reject the incoherent events by vetoing products from these nuclear breakups, such as, protons, neutrons, and photons~\cite{Chang:2021jnu, Tu:2020ymk, PhysRevD.111.072013, Chang:2022hkt, PhysRevC.104.065205}. As shown in the study of Ref.~\cite{PhysRevD.111.072013}, the unique capabilities offered from the second interaction region enable the discrimination of coherent and incoherent diffractive events across a wide range of momentum transfer. The purpose of this study is to see how many coherent events can be detected by tagging the intact nucleus, using the most up-to-date FF detector simulations of IR-8, which are based on the preliminary conceptual design from Ref.~\cite{PhysRevD.111.072013}. We use the eSTARlight event generator\footnote{\url{https://github.com/eic/estarlight}} ~\cite{Lomnitz:2018axr, Lomnitz:2018juf} to simulate coherent diffractive events, where each event consists of the produced VM at mid-rapidity, and an intact nucleus at forward-rapidities. The results will provide significant insight into the detector proposal and future improvements of IR-8 at the EIC. 

This paper is organized as follows. In Sec .~\ref {sec:IR8}, a general description of the proposed beam line for the second EIC detector at IR-8 will be presented. In Sec .~\ref {sec:detectors}, the forward detectors along the outgoing hadron beam will be described. In Sec .~\ref {sec:generator}, the event generator - eSTARlight will be briefly introduced. In Sec .~\ref {sec:result},  the results of the detection performance of coherent light nuclei will be shown, followed by a summary in Sec.~\ref{sec:summary}.

\section{\label{sec:IR8} Proposed IR-8 Layout}

The second interaction region~\cite{gamage:ipac2022-mopotk046} will be located at the eight o'clock position (IR-8) of the current RHIC complex at BNL. Compared to IR-6~\cite{ABDULKHALEK2022122447}, it will feature distinct blind spots in fiducial acceptance because of its 35 mrad crossing angle, which is larger than 25 mrad in
IR-6 and arises primarily from the geometric constraints of the existing experimental hall and tunnel. This larger angle introduces different pseudo-rapidity blind spots for IR-8, making it harder to achieve acceptance at high pseudo-rapidity ($\eta\sim$3.5) in the central detector. However, combining data from both experiments will improve overall EIC acceptance, exceeding what a single experiment could achieve. Furthermore, the current pre-conceptual design of IR-8 improves the acceptance of the forward detector for low transverse momentum ($p_{T}$) forward scattered particles, thereby strengthening the exclusive, tagging, and diffractive physics programs. Notably, the new hadron beam line design for the second EIC interaction region incorporates an optical configuration with a secondary focus positioned approximately 45$~$m downstream of interaction point, realized through the addition of dipole and quadrupole magnets. Analogous to the focus at the interaction point, this secondary focus reduces the transverse beam profile, enabling the detection of particles with minimal magnetic rigidity variation and small scattering angles near $\sim$0 mrad.

\section{\label{sec:detectors} Far-Forward Detectors}

\begin{figure}[tbh]
\includegraphics[width=\linewidth]{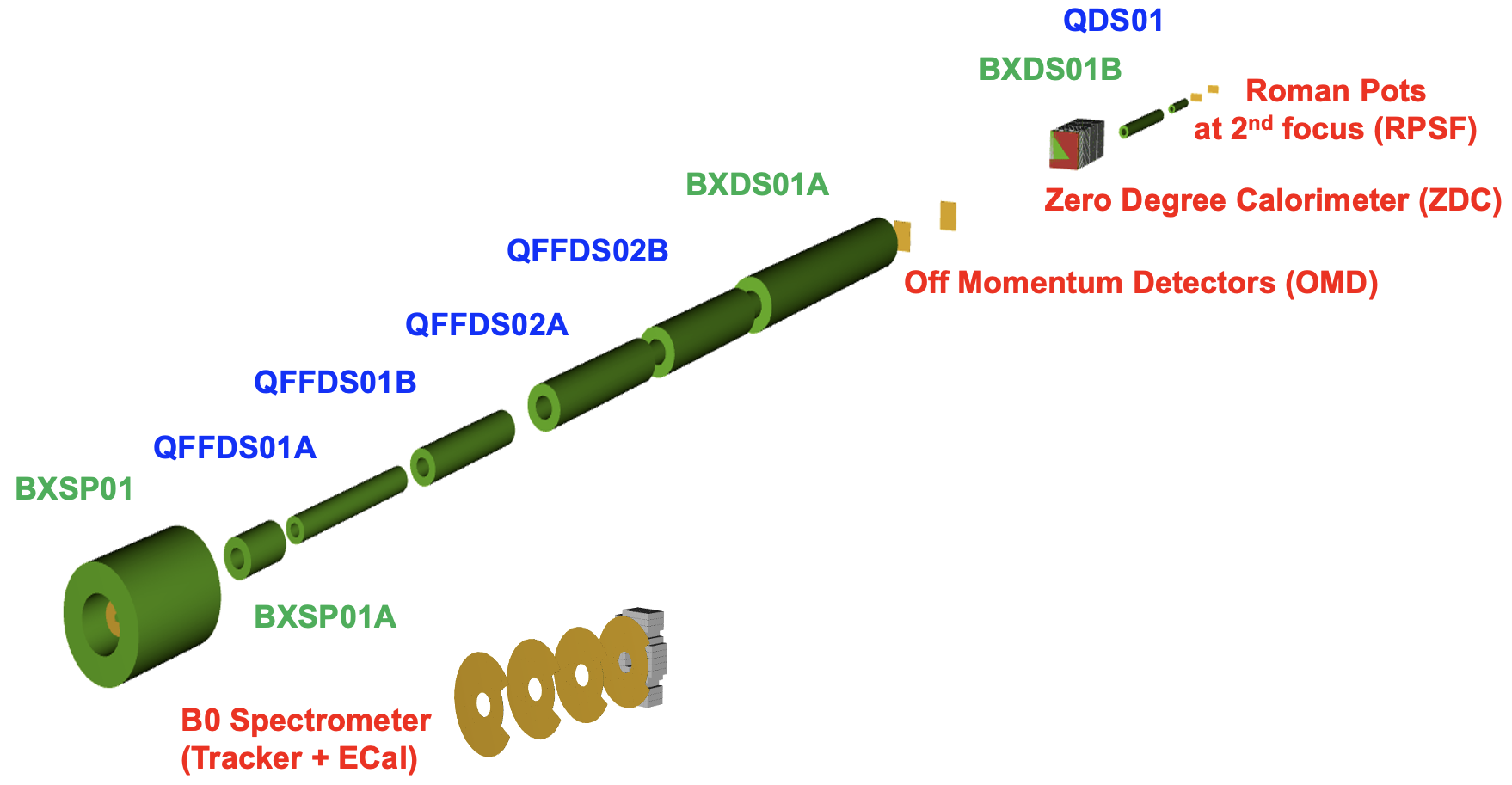}
  \caption{The layout of the far-forward IR design for the outgoing hadron beam direction at the second interaction region (IR-8) of the EIC. It includes four dipole (green), five quadrupole (blue) magnets in the ion beamline, and four far-forward detector subsystems. The figure is taken from ~\cite{PhysRevD.111.072013}.}
  \label{fig:ir8forward}
\end{figure}

\begin{table*}[thb]
%\fontsize{12}{15}\selectfont
\fontsize{11}{13}\selectfont
\begin{center}
\begin{tabular}{|c|c|c|c|c|}
\hline
  \textbf{Detectors} & $z$ Positions [m]  & $\theta$ [mrad] & Note\\
\hline
 B0 spectrometer & (5.8, 6.4) &  $5.0<\theta<20$ & tracker and calorimeter\\
 Off-momentum detector & (25.9, 28.0)   &  $0<\theta<5.0$ & 40\% to 60\% rigidity\\
 Zero-degree calorimeter & 40 & $\theta<5.5$ & full acceptance at $\theta<5$ mrad \\
 Roman Pots & (44.0, 45.5)  & $0<\theta<5.0$ & 10$\sigma$ cut \\
 
\hline
 \end{tabular}
 \end{center}
  \caption{\label{tab:table1} Summary of the physical $z$ positions, maximum polar angular acceptance for the far-forward detectors at IR-8 of the EIC~\cite{ABDULKHALEK2022122447, PhysRevD.111.072013}.}
 \end{table*}

To conduct measurements in diffractive physics, the EIC detector must encompass a wide acceptance range for both scattered charged and neutral particles, extending beyond the central detector to the far-forward region. Since many exclusive final states include an outgoing hadronic component essentially collinear with the hadron beam itself, it is crucial to incorporate specialized detectors within the interaction region.

To capitalize on new physics possibilities by utilizing this secondary focus at IR-8, we have adopted a preliminary conceptual design for the necessary far-forward detectors from Ref.~\cite{PhysRevD.111.072013}, aligning with the proposed beam line geometry and field configuration, as illustrated in Fig.~\ref{fig:ir8forward}. It includes the B0 spectrometer, the Off-Momentum Detector (OMD), a Zero-Degree Calorimeter (ZDC), and Roman Pots at the secondary focus (RPSF). The $z$ positions and angular acceptances for these far-forward detectors are detailed in Table.~\ref{tab:table1}. Some general considerations used to establish baseline particle acceptances and detector resolutions for this study are fully simulated using Eicroot~\cite{ref:EICROOT} and GEANT~\cite{Brun:1987ma, GEANT4}.

Starting from the interaction point and moving downstream, the first detector subsystem is the B0 spectrometer \footnote{The name ``B0" comes from the name of the magnet in IR-6, which houses the same style of spectrometer for the ePIC detector. The B0 name has been used here to make it easier to see the complementary detector concept shared between the two IRs.}. The B0 spectrometer is designed to detect scattered protons and photons within the angular range of 5 mrad to 20 mrad, which is larger than that of the far-forward region. It comprises four silicon tracking layers and a 10 cm lead-tungsten crystal calorimeter. This entire assembly is enclosed within the 1.3~T BXSP01 dipole magnet, designated as "B0" in accordance with the IR-6 configuration. The four tracking layers, in conjunction with the magnetic field, enable full momentum reconstruction of charged particles via standard tracking methodologies.

The next detector subsystem is the off-momentum detector (OMD). This subsystem comprises two silicon planes, spaced two meters apart, and placed just after the BXDS01A dipole magnet outside the beam pipe vacuum. The OMD is engineered to detect scattered protons and charged particles originating from nuclear dissociation, with rigidities below 60\% of those of the beam particles. The lower magnetic rigidity causes these charged particles to experience a larger bending angle in the dipole magnets, causing them to be steered out of the beam pipe. 

The third one is the Zero-Degree Calorimeter (ZDC), positioned immediately upstream of the BXDS01B dipole magnet. The ZDC is built to detect high-energy neutrons and photons, as well as low-energy photons ($>$100 MeV), critical for distinguishing between incoherent and coherent events in scatterings with heavy nuclei~\cite{PhysRevD.111.072013}. This study focuses on the imaging probability of coherent light nuclei at the second interaction region. Given that the ZDC is dedicated to detecting neutral particles, it has been excluded from the present analysis.

The final detector subsystem is the Roman Pots at the secondary focus (RPSF). The RPSF is designed to detect scattered protons and nuclear fragments with minimal rigidity changes, up to a scattering angle of 5 mrad. It is equipped with two silicon tracking layers at the secondary focus to observe scattered particles with $p_{T} \sim 0$. However, the detector must maintain a safe distance from the beam—one typically set by the ``10$\sigma$ rule", where $\sigma$ denotes the transverse beam size and which is calculated by:

\begin{equation}
     \sigma_{x,y} = \sqrt{\epsilon_{x,y}\beta(z)_{x,y} + (D_{x,y}\frac{\Delta p}{p})^{2}}.
     \label{eq:beamsize}
 \end{equation}

\noindent Here $\epsilon$ is the beam's emittance, $\beta$ is the beta-function, $D$ is the momentum dispersion, and $\Delta p/p$ is the relative momentum spread.

For the beam parameters outlined in the EIC Conceptual Design Report~\cite{ref:EICCDR}, 1$\sigma$ is 0.15 millimeters in the $x$ direction and 0.1 millimeters in the $y$ direction for the top energy ion beams. Different beam optics configurations allow for adjustment of the transverse beam size at the RP detectors, introducing a trade-off between detector acceptance and total luminosity.

\section{\label{sec:generator} Event generator}

To evaluate the impact on the tagging efficiency and acceptance (hereafter, we use detection efficiency or efficiency for simplicity unless specified otherwise) based on the pre-conceptual design of IR-8, including the secondary focus and the far-forward detectors, we used the eSTARlight event generator to generate coherent events of vector meson production: $e+\rm{A} \rightarrow \it{e'}+\rm{A}'+\rm{VM}$. eSTARlight is a Monte Carlo that simulates coherent vector meson photoproduction and electroproduction in electron-ion collisions. This is similar to the STARLight event generator~\cite{Klein:2016yzr} in terms of its physics model, which has been extensively used and applied in the ultra-peripheral collisions in heavy ion physics. It is capable of generating a diverse array of final states across varying center-of-mass energies for different collision systems, with the photon virtuality set at arbitrary values, but no saturation effect is included~\cite{Lomnitz:2018axr, Lomnitz:2018juf}.

For this study, we ran the eSTARlight event generator to produce various samples of $e+\rm{A}\rightarrow \it{e'}+\rm{A}'+\rm{VM}$ ($\rm{A= ~^{2}\rm{D}},~^{3}\rm{He},~^{4}\rm{He}, ~^{7}\rm{Li},~^{9}\rm{Be},~^{12}\rm{C}, ~^{16}\rm{O}$ and $\rm{VM} = \it{J}/\psi$, $\phi$, $\rho$), across different energies for both electron and nucleus beams. Each sample, consisting of 10 million events with a kinematic range of $0.1<Q^{2}<100~\mathrm{GeV}^{2}$, was passed through the EIC afterburner~\cite{afterburner_github}. This step accurately incorporates the crossing angle and beam effects, such as angular divergence and momentum spread, ensuring the simulation's correct collision frame and smearing effects induced by the beam. These factors are particularly crucial as forward particle acceptance and momentum resolution are highly sensitive to them. The vast majority of scattered intact nuclei are tagged by the RPSF. Only less than $1\%$ of nuclei, those with lower collision energies($10\times100~\rm{GeV}^{2}$, and $5\times41~\rm{GeV}^{2}$), can additionally be identified via the OMD. The current simulation only accounts for the acceptance effect and does not incorporate the efficiencies of the detector. Additionally, we did not account for the efficiency and acceptance of the reconstructed distribution, which is to showcase our detection performance at its detector level form.

\section{\label{sec:result} Results}
In this section, we investigate the detection performance of coherent light nuclei at the second interaction region at the EIC. We will first present the detection efficiency of various coherent light nuclei, ranging from deuterium ($^{2}$D) to oxygen ($^{16}$O) at their top energies, where the electron energy is 18 GeV and the energy of various nuclei is calculated by $Z/A\times275~\rm{GeV}$($Z$ is the proton number and $A$ is the atomic number), as a function of the invariant mass of the produced hadronic system $W$. At the same time, the correlation of squared momentum transfer $Q^{2}$ and parton momentum fractions $x$ for detection efficiency for $e$$^{2}$D, $e$$^{7}$Li, and $e$$^{16}$O with the top collision energies are shown. Following that, we will compare the detection efficiency of coherent $^{3}$He nucleus across different collision energies. After that, the results of coherent $^{3}$He nucleus with different VMs production will be demonstrated. We also talk about the momentum transfer $|t|$ distributions in the MC-generated level and detected by far-forward detectors at IR-8 for $e$$^{3}$He and $e$$^{7}$Li collisions with top energies. The resulting Fourier transforms from $|t|$ distributions are shown as well to showcase the physics potential of the imaging program at the second interaction region.

\subsection{\label{subsec:DifferentA} The detection efficiency of various nuclear species}

\begin{figure}[tbh]
\includegraphics[width=\linewidth]{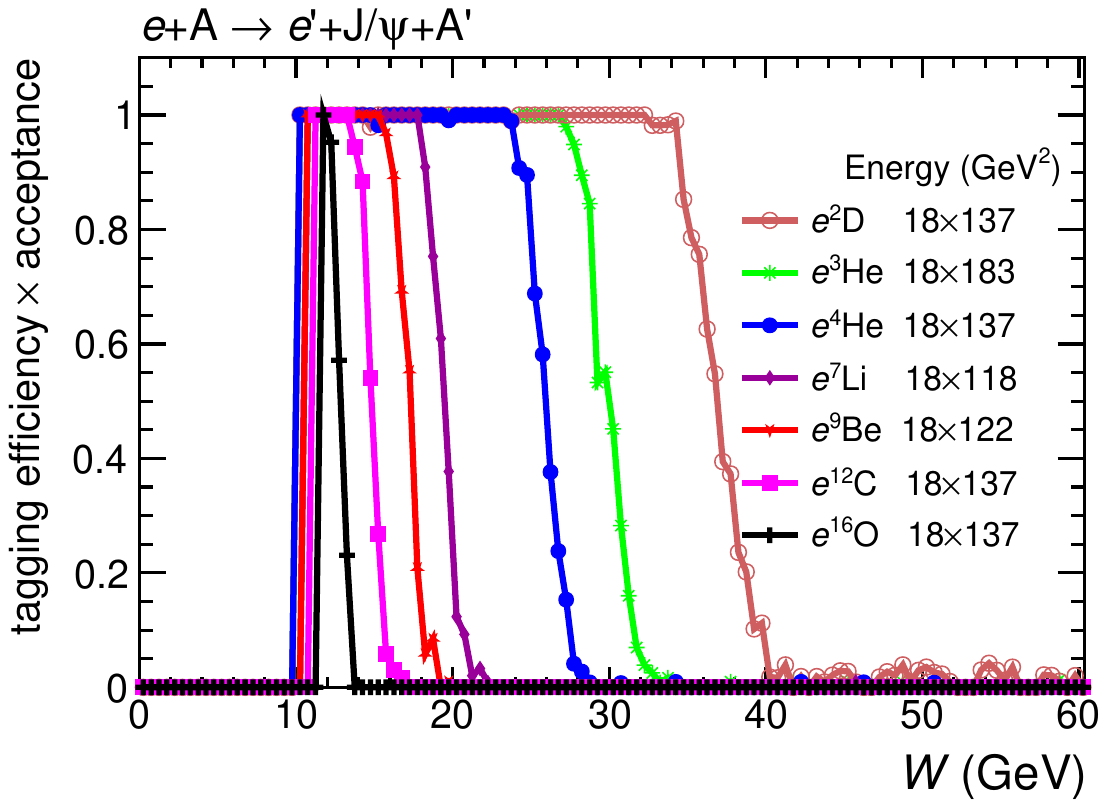}
  \caption{The detection efficiency as a function of $W$ for coherent $J/\psi$ production in various light nuclei. The results for different coherent nuclei are shown with different colors.}
  \label{fig:W_diffA}
\end{figure}

\begin{figure*}[tbh]
\centering
\includegraphics[width=0.3\textwidth]{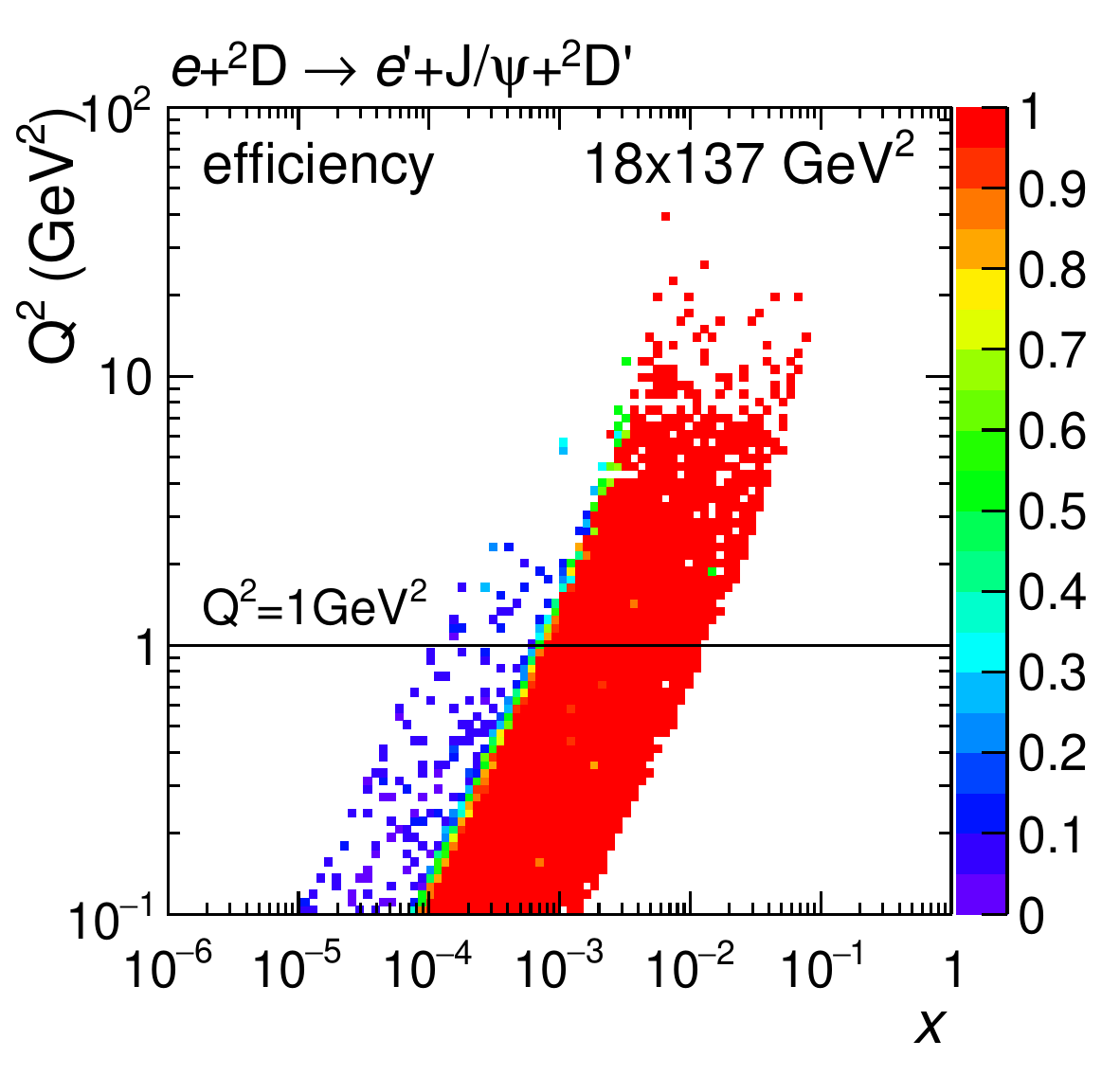}
\includegraphics[width=0.3\textwidth]{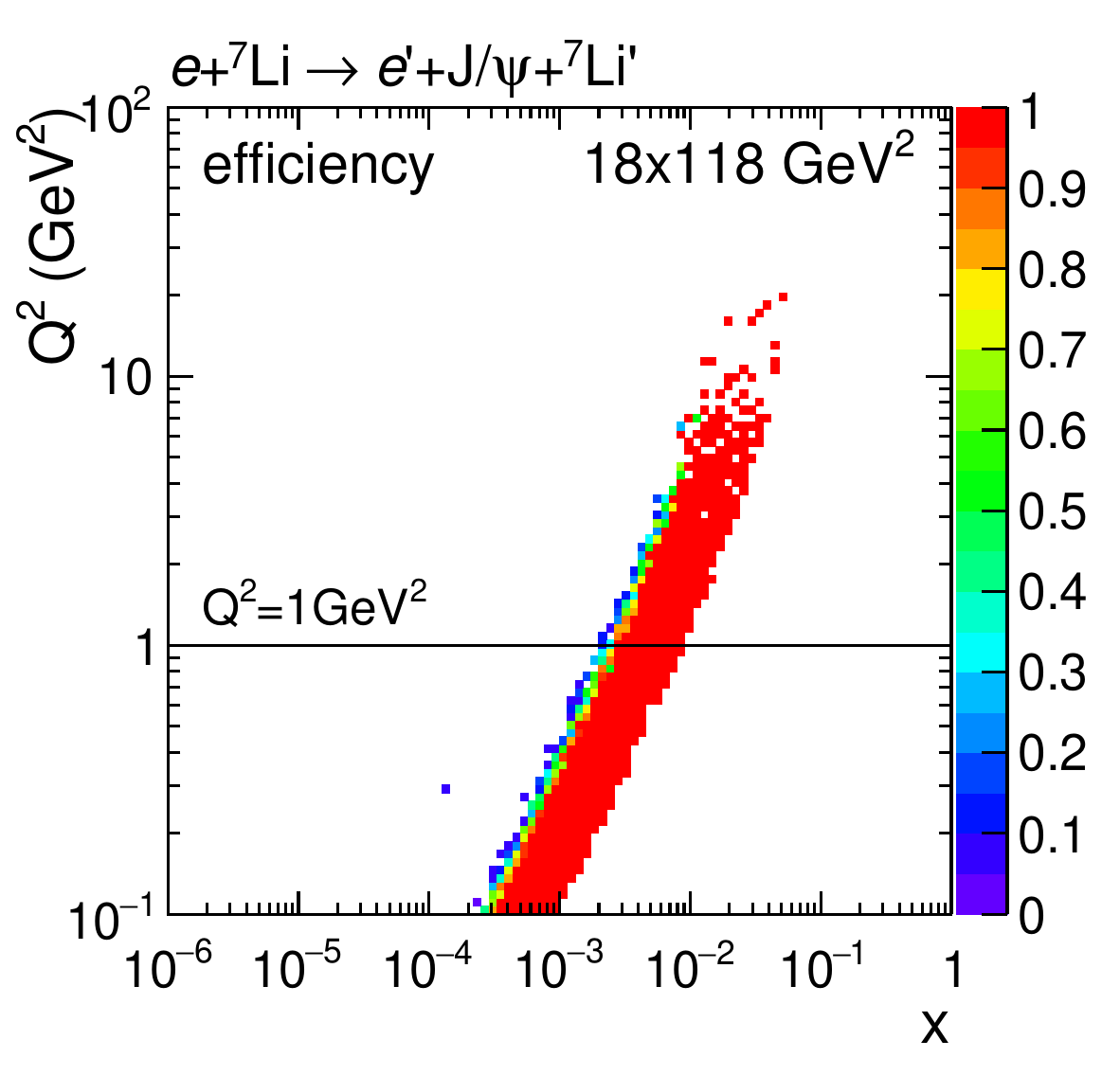}
\includegraphics[width=0.3\textwidth]{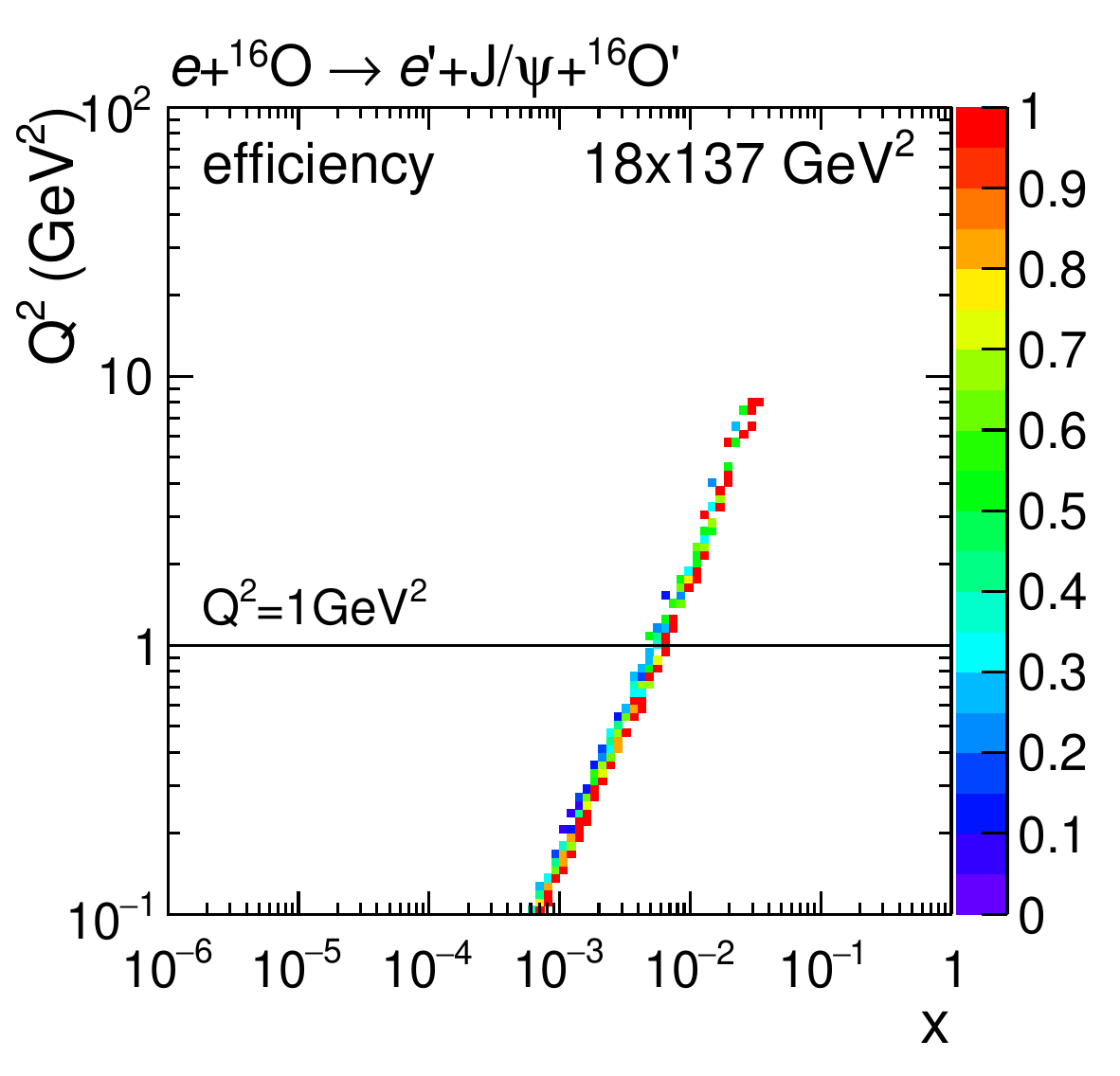}
\caption{The correlation of $Q^{2}$ and $x$ for detection efficiency for coherent $J/\psi$ production in $e$$^{2}$D, $e$$^{7}$Li, and $e$$^{16}$O collisions with the top collision energies.}
\label{fig:eA_Q2vsX}
\end{figure*}

\begin{figure*}[tbh]
\centering
\includegraphics[width=0.3\textwidth]{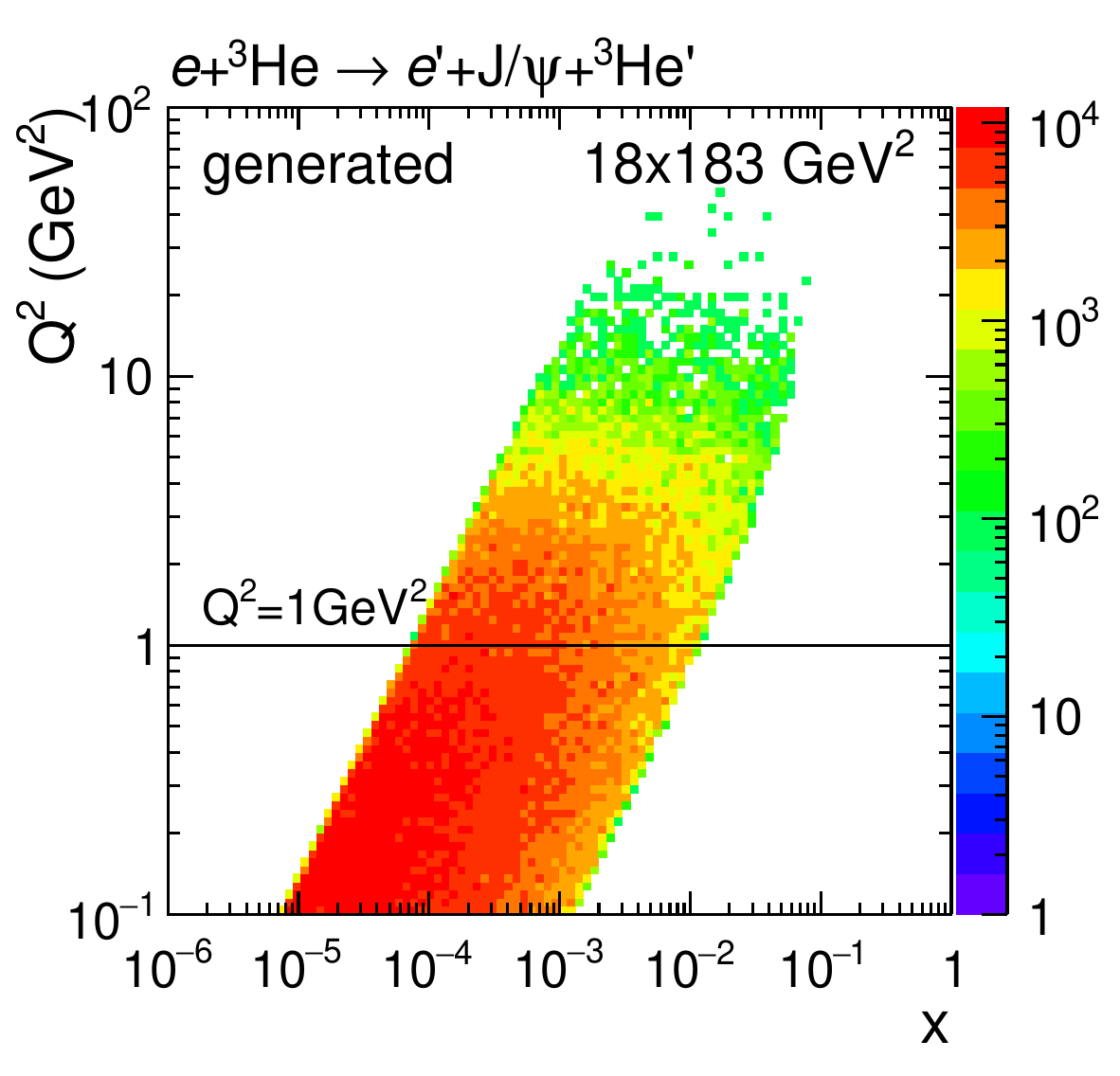}
\includegraphics[width=0.3\textwidth]{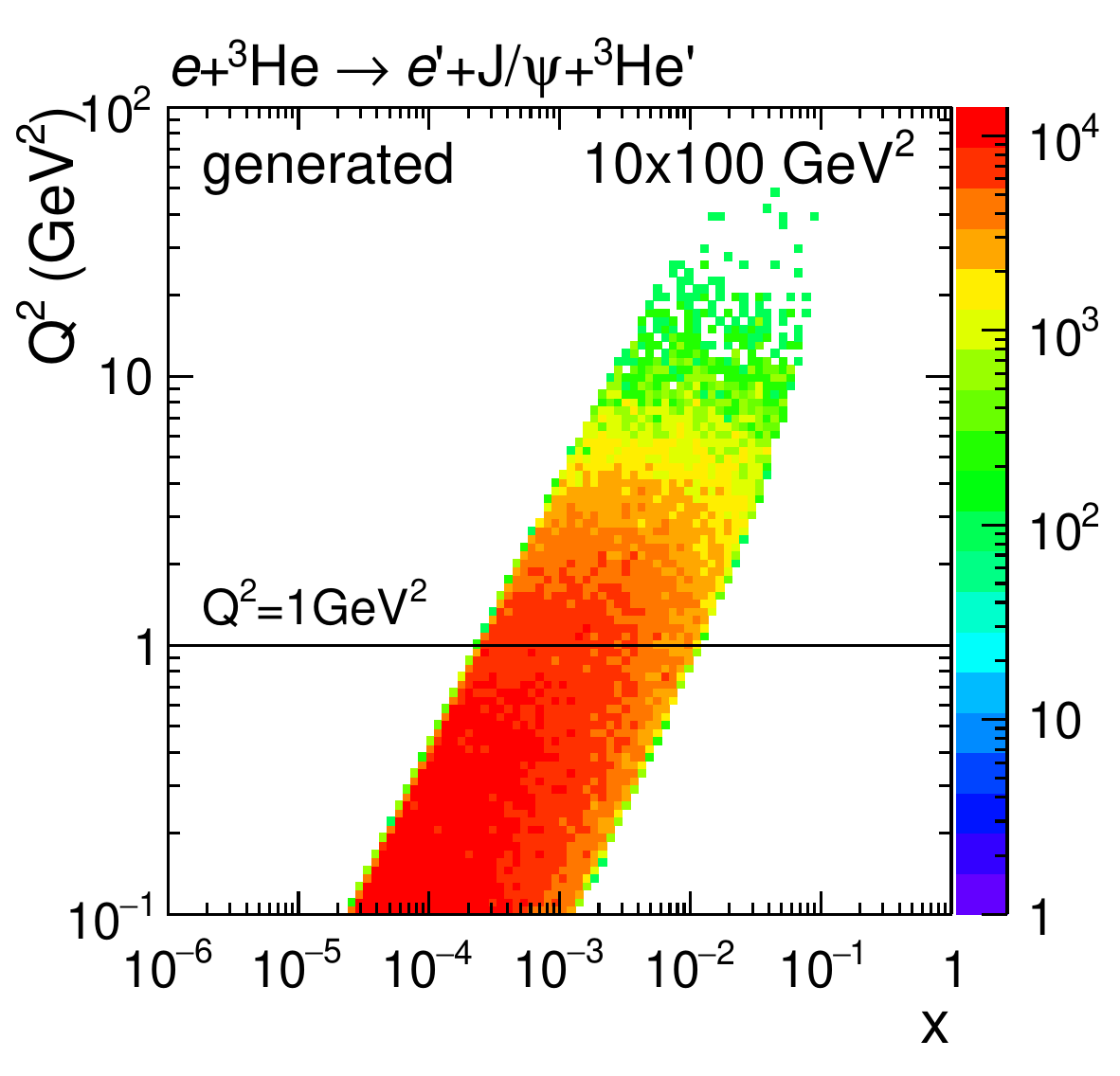}
\includegraphics[width=0.3\textwidth]{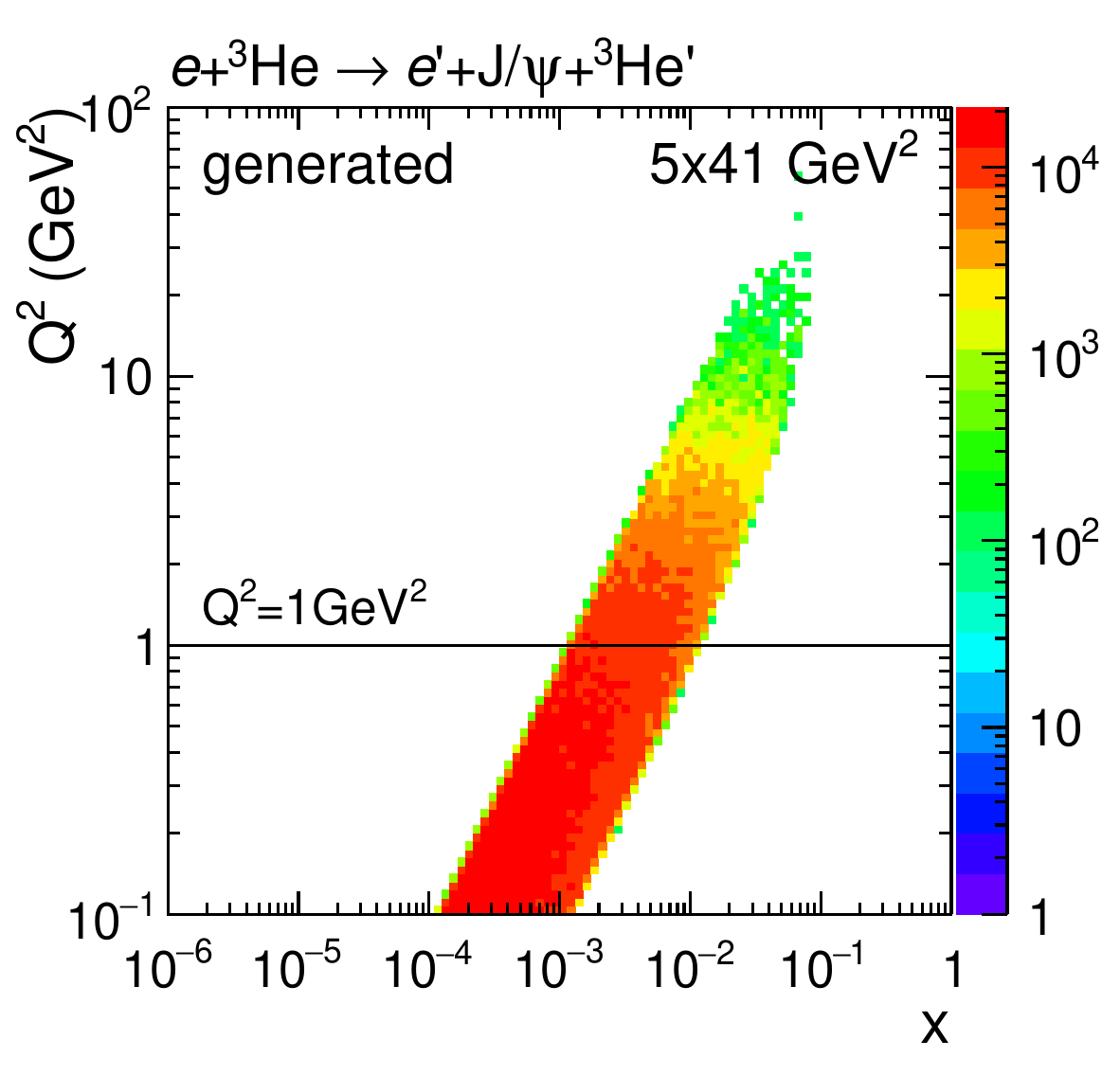}
\includegraphics[width=0.3\textwidth]{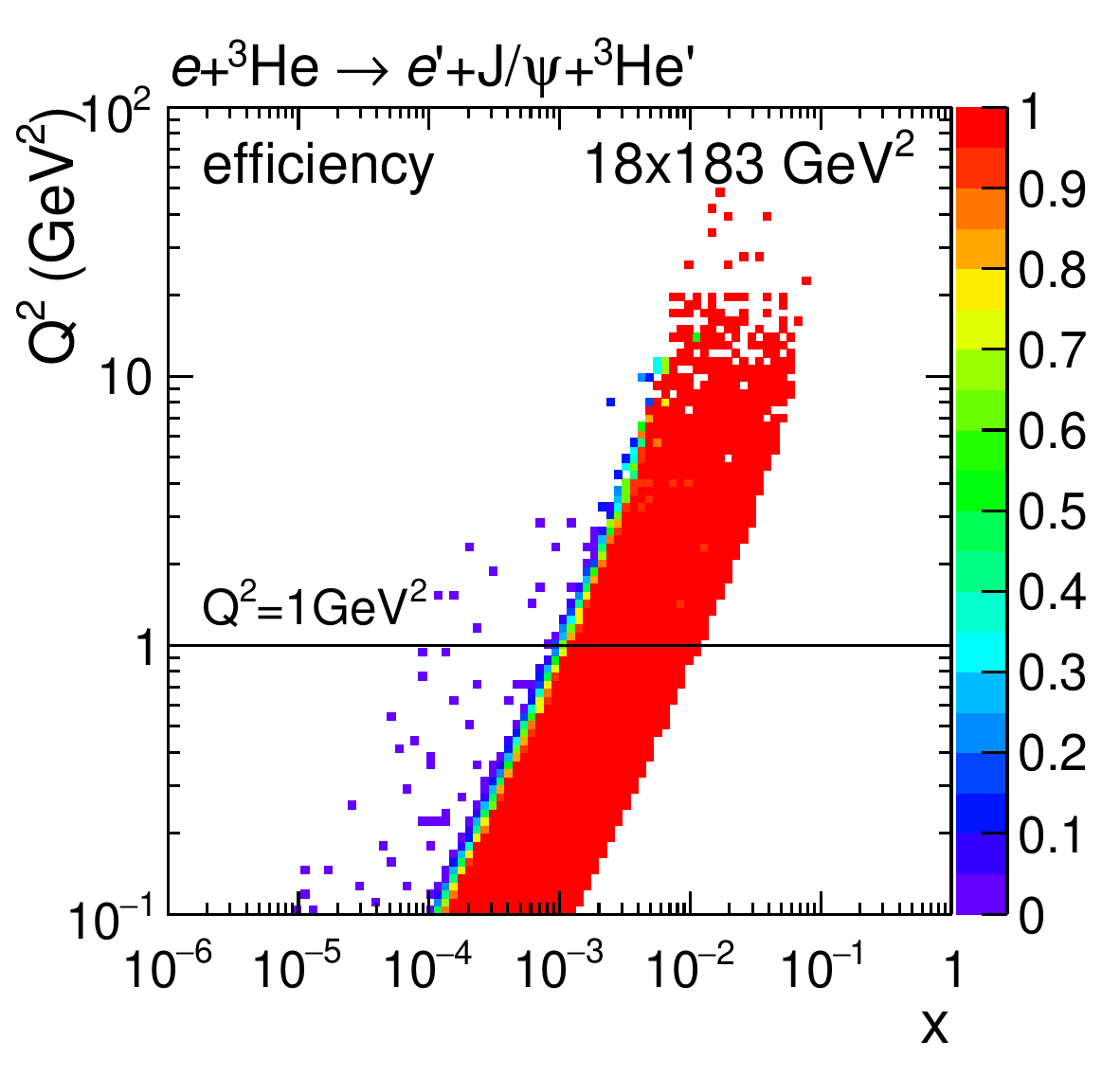}
\includegraphics[width=0.3\textwidth]{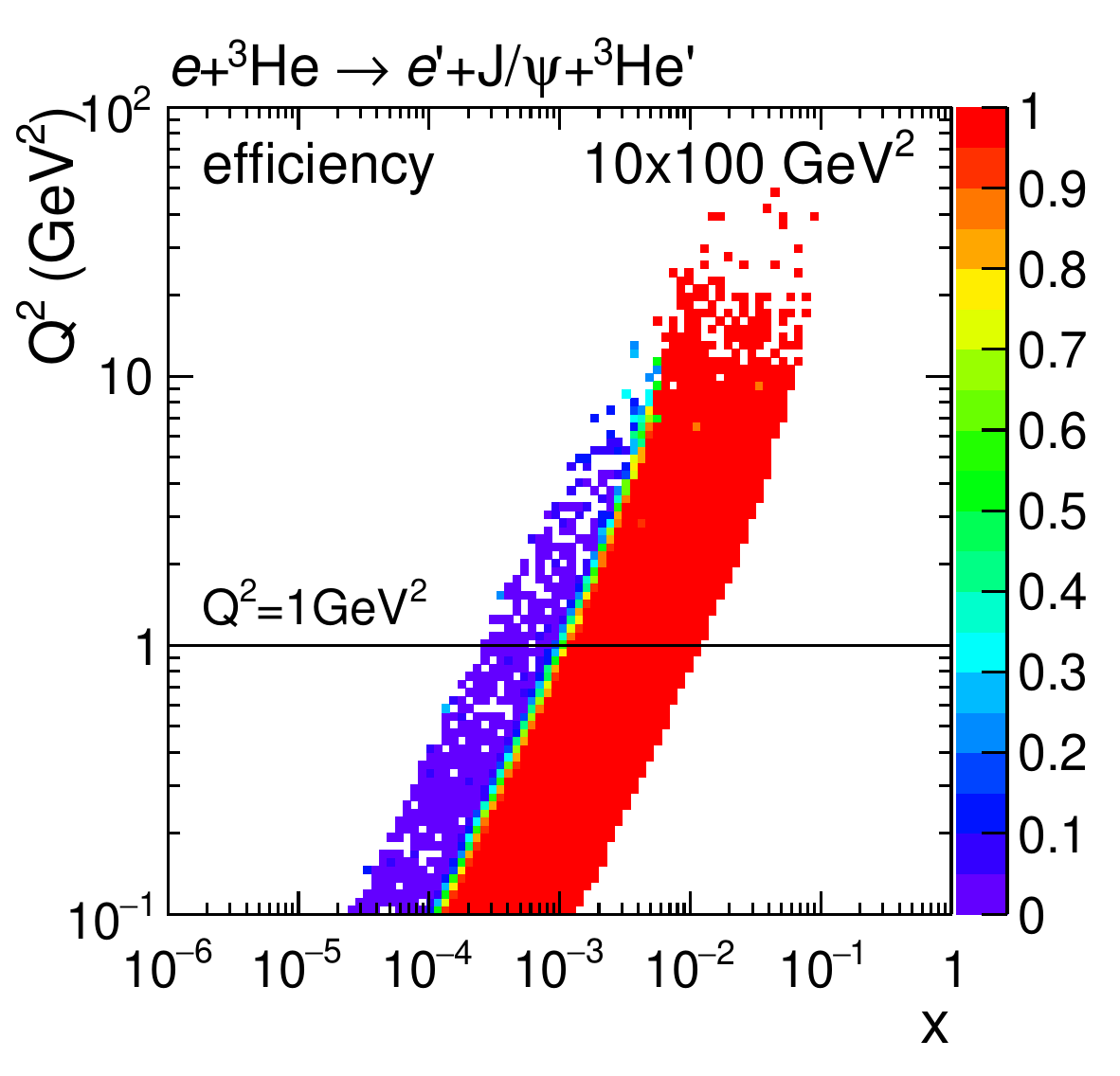}
\includegraphics[width=0.3\textwidth]{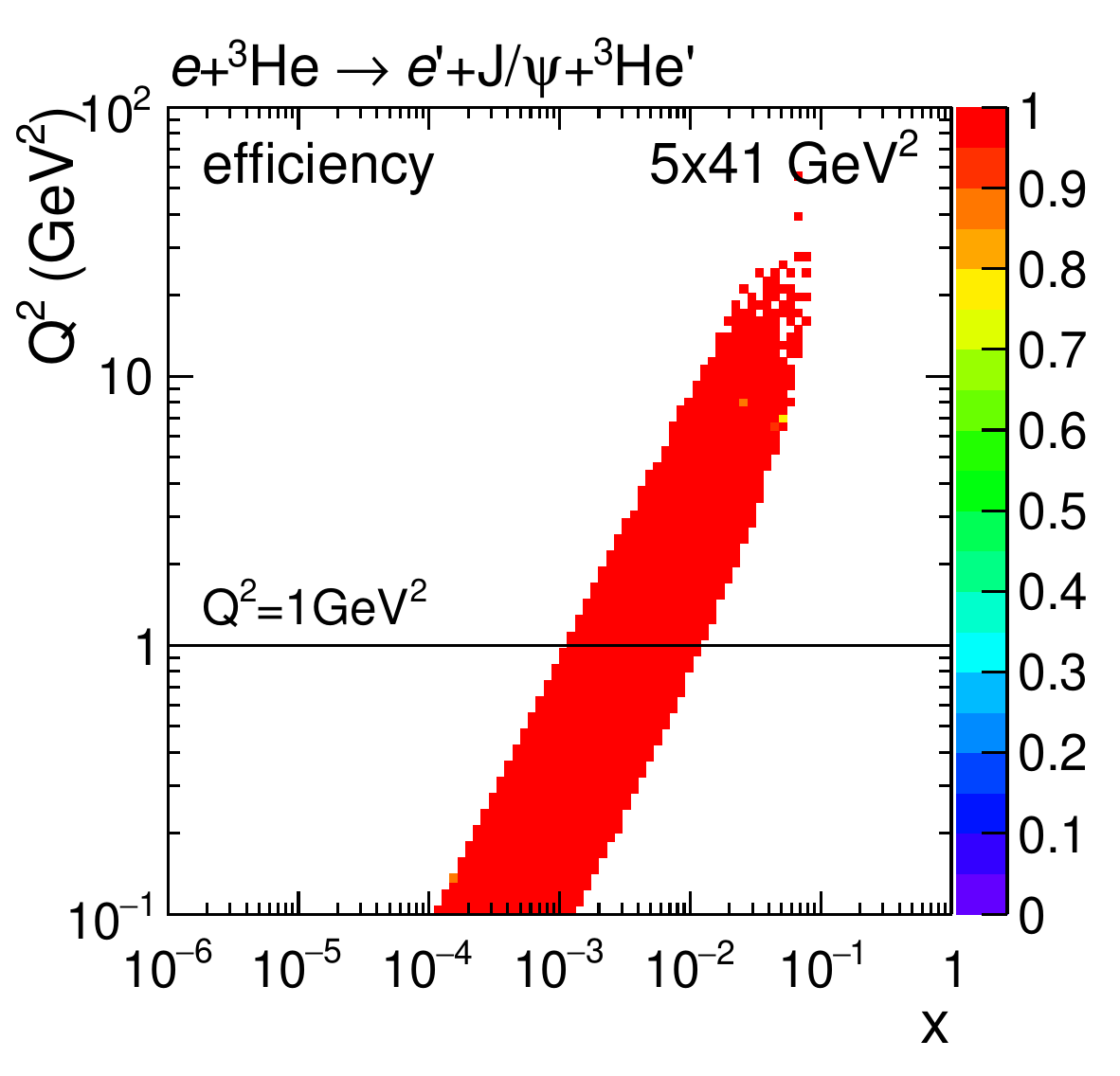}
%\caption{The correlation of $Q^{2}$ and $x$ for tagging efficiency $\times$ acceptance for coherent $J/\psi$ production in $e$$^{3}$He collisions with different energies: $18\times183~\rm{GeV}^{2}$, $10\times100~\rm{GeV}^{2}$, and $5\times41~\rm{GeV}^{2}$.}
%\label{fig:eHe_Q2vsX}
\caption{The correlation of $Q^{2}$ and $x$ for coherent $J/\psi$ production in $e$$^{3}$He collisions with different energies: $18\times183~\rm{GeV}^{2}$, $10\times100~\rm{GeV}^{2}$, and $5\times41~\rm{GeV}^{2}$. Top row: the distributions for MC-generated events by eSTARlight. Bottom row: the distributions for detection efficiency of the far-forward detectors of IR-8 at EIC.}
\label{fig:eHe_Q2vsX}
\end{figure*}

The detection efficiency for coherent $J/\psi$ production in various light nuclei as a function of $W$ is shown in Fig.~\ref{fig:W_diffA}. $W$ is given by 
\begin{equation}
     W=\sqrt{(P+q)^{2}},
     \label{eq:W}
 \end{equation}
where $P$ is the four momentum of the target nucleon, $q$ denotes the momentum transfer between the incoming and outgoing electron. 

These measurements were carried out with the top energies allowed in the EIC accelerator. The lighter nucleus allows for the detection of events with larger $W$. This is because the lighter nuclei, for the same value of $W$, exhibit a lower magnetic rigidity, which results in a greater bending angle within the dipole magnets. Consequently, these particles are more easily steered out of the beam pipe. The detectable $W$ value for ${e}^{2}$D collision events can go up to 40 GeV. However, the maximum detectable $W$ value for coherent $\it{e}$$^{16}$O events is 12 GeV. The global detection efficiency as the increasing nucleon number, from deuterium($^{2}$D) to oxygen($^{16}$O), is 47.12$\%$, 32.23$\%$, 29.42$\%$, 17.75$\%$, 12.37$\%$, 6.36$\%$, 1.59$\%$, respectively. 

In contrast to IR-6, the secondary focus at IR-8 significantly improves the tagging performance on light ions. For instance, the detection of coherent $J/\psi$ production in $e$$^{3}$He collisions with top energy is limited to $^{3}$He with $p_{T} > 200$ MeV/c at IR-6 (assuming a similar high-acceptance optics configuration for e+p), while the secondary focus in IR-8 enables tagging of $^{3}$He nuclei for $p_{T} \sim 0$. It should be noted that the optics assumption for IR-6 has a major impact on the potential low-$p_{T}$ acceptance improvement in IR-8 - as $A$ of the nucleus is increased, the impact of the secondary focus in IR-8 will become more pronounced, especially for higher $A$ where IR-6 optics essentially do not allow for any acceptance at low-$p_{T}$. 

It is important to note that, for IR-6, the low-$p_{T}$ acceptance can be further improved by tuning the beam parameters, such as the $\beta$ functions, which affect the low-$p_{T}$ cutoff via the safe distance of approach for the Roman pots. However, this comes with a significant trade-off in luminosity in IR-6, which is particularly important for exclusive measurements. Therefore, under the same or luminosity-optimized beam configurations, IR-8 offers a substantial advantage in Far-Forward tagging acceptance for light ions, especially at low-$p_{T}$. 

Fig.~\ref{fig:eA_Q2vsX} shows the correlation of squared momentum transfer $Q^{2}$ and parton momentum fractions $x$ for detection efficiency for $e$$^{2}$D, $e$$^{7}$Li, and $e$$^{16}$O with the top collision energy. The black lines drawn in the pictures are with $Q^{2} = 1~\rm{GeV}^{2}$. For $Q^{2} > 1~\rm{GeV}^{2}$ , the $x$ range for accepting events is $x>10^{-4}$ for $e$$^{2}$D collision events, $x>2\times10^{-3}$ for $e$$^{7}$Li collision events, and $x>4\times10^{-3}$ for $e$$^{16}$O collision events.

    \subsection{\label{subsec:DifferentE} The detection efficiency of different collision energies}

The correlation of $Q^{2}$ and $x$ for coherent $J/\psi$ production of $e$$^{3}$He collisions with different energies: $18\times183~\rm{GeV}^{2}$, $10\times100~\rm{GeV}^{2}$, and $5\times41~\rm{GeV}^{2}$, are shown in Fig.~\ref{fig:eHe_Q2vsX}. The top three plots show the distributions of MC-generated events by eSTARlight. The higher the center-of-mass energy, the lower the value of $x$ can be. At the top collision energy, the minimum value of $x$ can reach $7\times10^{-5}$ at $Q^2 \approx 1~\mathrm{GeV^{2}}$. However, for coherent photoproduction in the eSTARlight event generator, the maximum value of $x$ is closely related to the nuclear radius, as the maximum momentum transfer is constrained by the coherence condition, requiring the coherence length to be greater than the nuclear radius. This maximum momentum transfer is determined by the size of the nucleus; therefore, the maximum value of $x$ is independent of the center-of-mass energy of the collisions, remaining constant across different center-of-mass energies\cite{Lomnitz:2018juf}. 

The three plots in the bottom row depict the distributions of the corresponding detection efficiency. In scenarios with a fixed collision energy, the detection efficiency in the high $x$ range is much greater than that in the low $x$ range. Moreover, the detection efficiency does not show a significant dependence on $Q^{2}$. At the top collision energy, $32.23\%$ of the scattered $^{3}\rm{He}$ nuclei occur within a safe distance from the beam, allowing them to be distinguishable from the beam and detected by RPSF. At the energy of $10\times100~\rm{GeV}^{2}$, the total detection efficiency is $54.38\%$. Of this, $53.47\%$ are detected by the RPSF, with the remaining $0.91\%$ can be identified by the OMD. For the lowest collision energy, 5 GeV electron beams on 41 GeV $^{3}\rm{He}$ beams, the detection efficiency is $99.77\%$. RPSF detects $99.41\%$ of the coherent $^{3}$He nuclei, whereas OMD is employed to identify the remaining ones.

%Furthermore, we also explored things in different $W$ ranges. For a fixed collision energy, an increase in the value of $W$ corresponds to a decrease in the minimum achievable value of $x$. Specifically, for values of $W$ less than 20 GeV, the lowest $x$ can be is approximately $2.5 \times 10^{-6}$. Conversely, within the range of 50 GeV to 100 GeV, this minimum value of $x$ decreases to around $10^{-7}$. The detection efficiency distributions of $Q^{2}$ vs. $x_{Bj}$ for the different $W$ bins are shown in the bottom row. The detection efficiencies in these three $W$ intervals, ranging from low to high, are $99.94\%$, $37.22\%$, and $0.07\%$, respectively.

    \subsection{\label{subsec:DifferentVM} The detection efficiency for events with different vector mesons}

\begin{figure}[tbh]
\includegraphics[width=\linewidth]{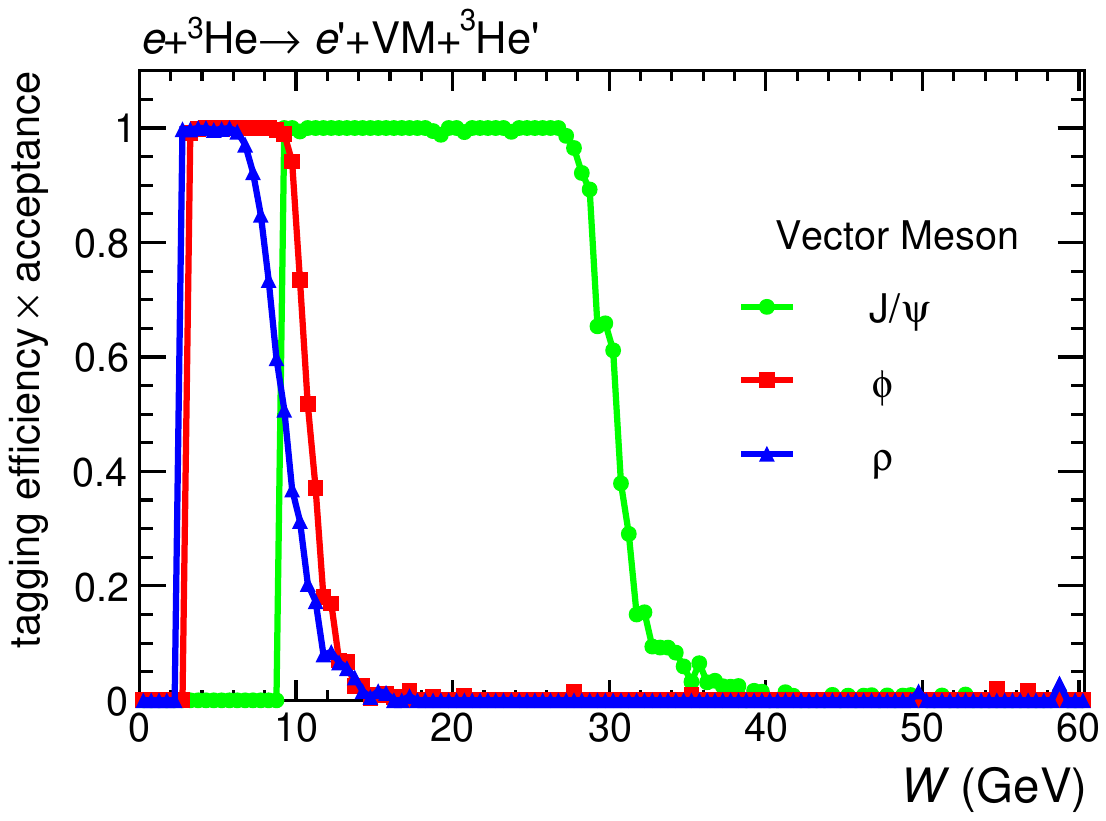}
  \caption{The detection efficiency as a function of $W$ for different coherent VM productions of $e^{3}\rm{He}$ with collision energy of $18\times183~\rm{GeV}^{2}$. The results with different VM productions are represented by different colors.}
  \label{fig:W_diffVM}
\end{figure}

% For a given gluon density distribution, using different VM probes may yield differing coherent cross sections due to their dipole sizes, thereby giving different experimental constraints to the underlying gluon density distributions
In exclusive vector meson production, one can compare different vector meson species as a probe to different equivalent dipole sizes that are determined by the mass of the dipole quark-antiquark pairs. Differences found in the momentum transfer $t$ distributions could be a sensitive probe of the saturation dynamics~\cite{Toll:2012mb, Klein:2016yzr, Sambasivam:2019gdd, PhysRevC.81.025203}. In Fig.~\ref{fig:W_diffVM}, the detection efficiency as a function of $W$ for different coherent VMs production ($J/\psi$, $\phi$ and $\rho$) of $e^{3}\rm{He}$ with collision energy of $18\times183~\rm{GeV}^{2}$ is shown. The results for coherent events with different VM productions are represented by different colors. Since the mass of the $\rho$ particle is the lightest, the distribution of coherent $\rho$ production goes to the smallest $W$ value, which is $2~\rm{GeV}$. Moreover, within the range of $W < 6~\rm{GeV}$, its detection efficiency is nearly 1. However, once $W$ surpasses $6~\rm{GeV}$, the detection efficiency decreases rapidly, and at around $W = 14~\rm{GeV}$, the detection efficiency drops to 0. In the case of coherent $\phi$ production, the detection efficiency remains nearly 1 for the $W$ range of $3~\rm{GeV}$ to $9~\rm{GeV}$, and then it rapidly decreases to nearly 0. For coherent $J/\psi$ production, the detectable range of $W$ is relatively broad, with detection efficiency close to 1 in the range of $10~\rm{GeV}<\it{W} < \rm{27~GeV}$, and it drops to 0 at $W\sim 37~\rm{GeV}$.

\subsection{\label{subsec:tDis} Momentum transfer $t$ distribution}

\begin{figure}[tbh]
\includegraphics[width=0.4\textwidth]{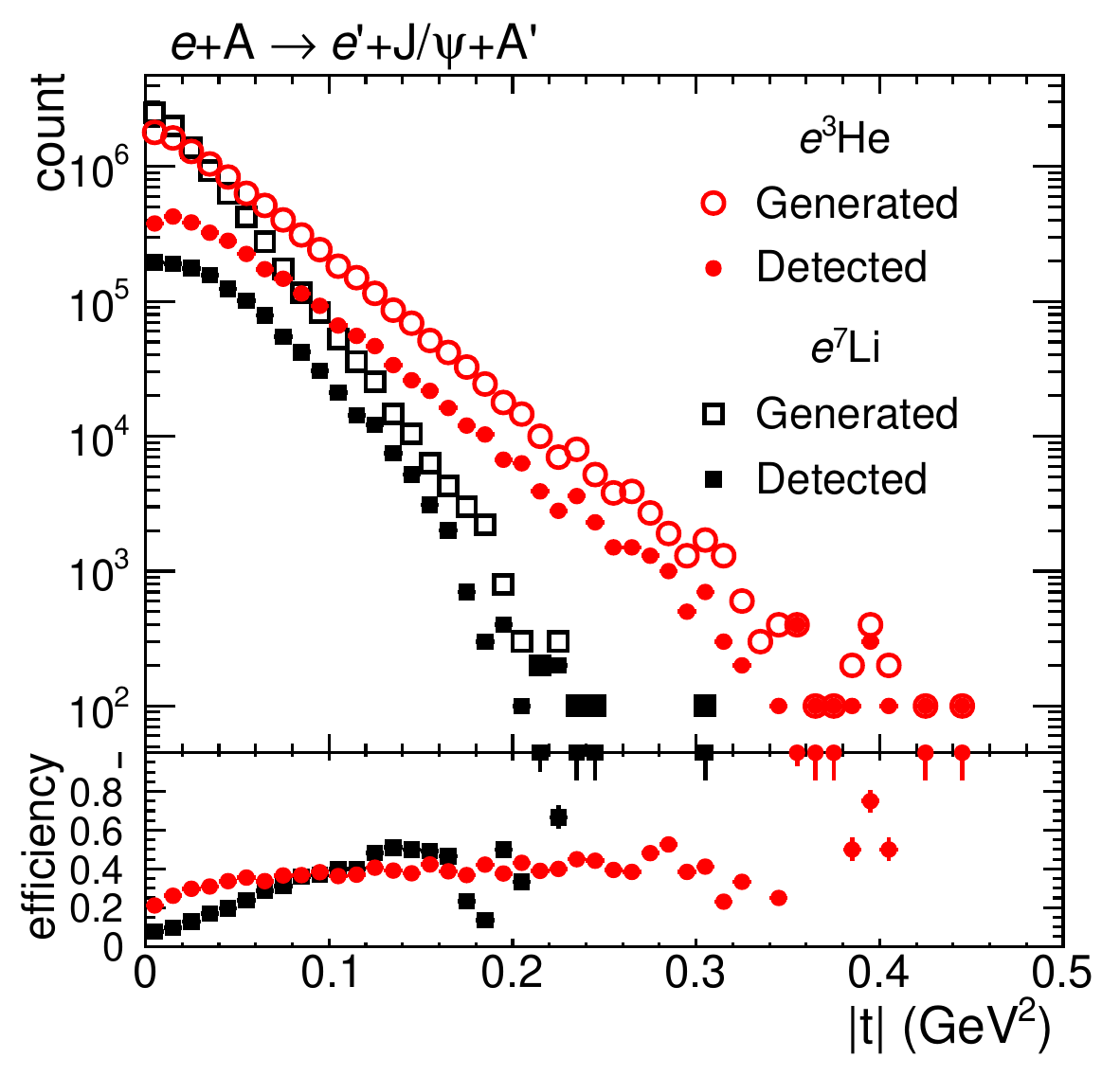}
  \caption{The momentum transfer $|t|$ distributions in the MC-generated level and detected by far-forward detectors at IR-8 for coherent $J/\psi$ production in $e^{3}$He collisions and $e^{7}$Li collisions at their respective top energies. At the bottom of the figure are the tagging efficiency distributions.}
  \label{fig:t}
\end{figure}

%\begin{figure}[tbh]
%\includegraphics[width=0.8\linewidth]{figures/Plots_eHe3/Efficiency_Wvst_eHe3_275.pdf}
% \caption{\label{fig:Wvst}The correlation of $W$ and $|t|$ for tagging efficiency $\times$ acceptance for $e^{3}$He collisions with energy of $18\times183~\rm{GeV}^{2}$.}
%\end{figure}

\begin{figure*}[tbh]
\centering
\includegraphics[width=0.4\textwidth]{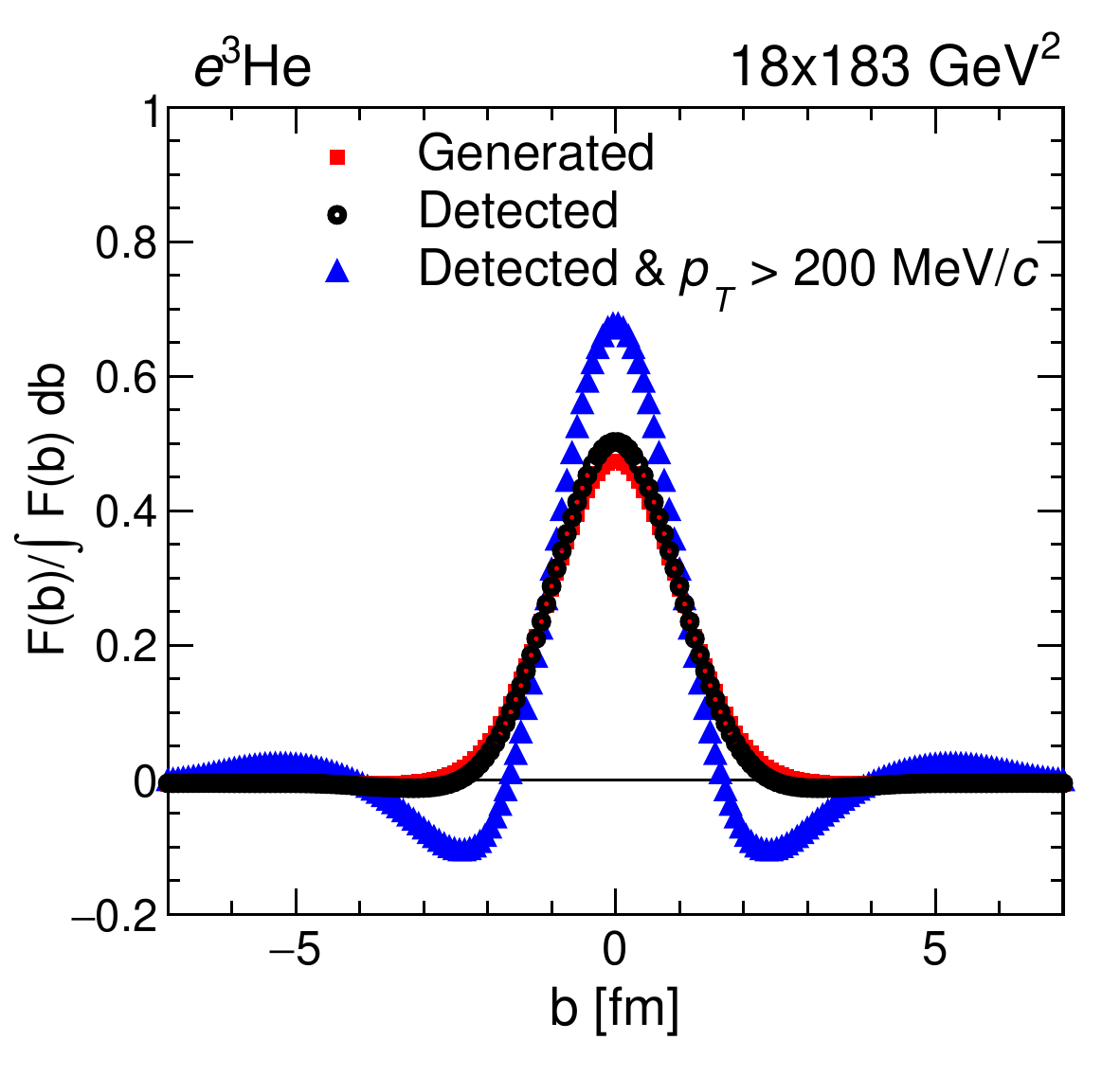}
\includegraphics[width=0.4\textwidth]{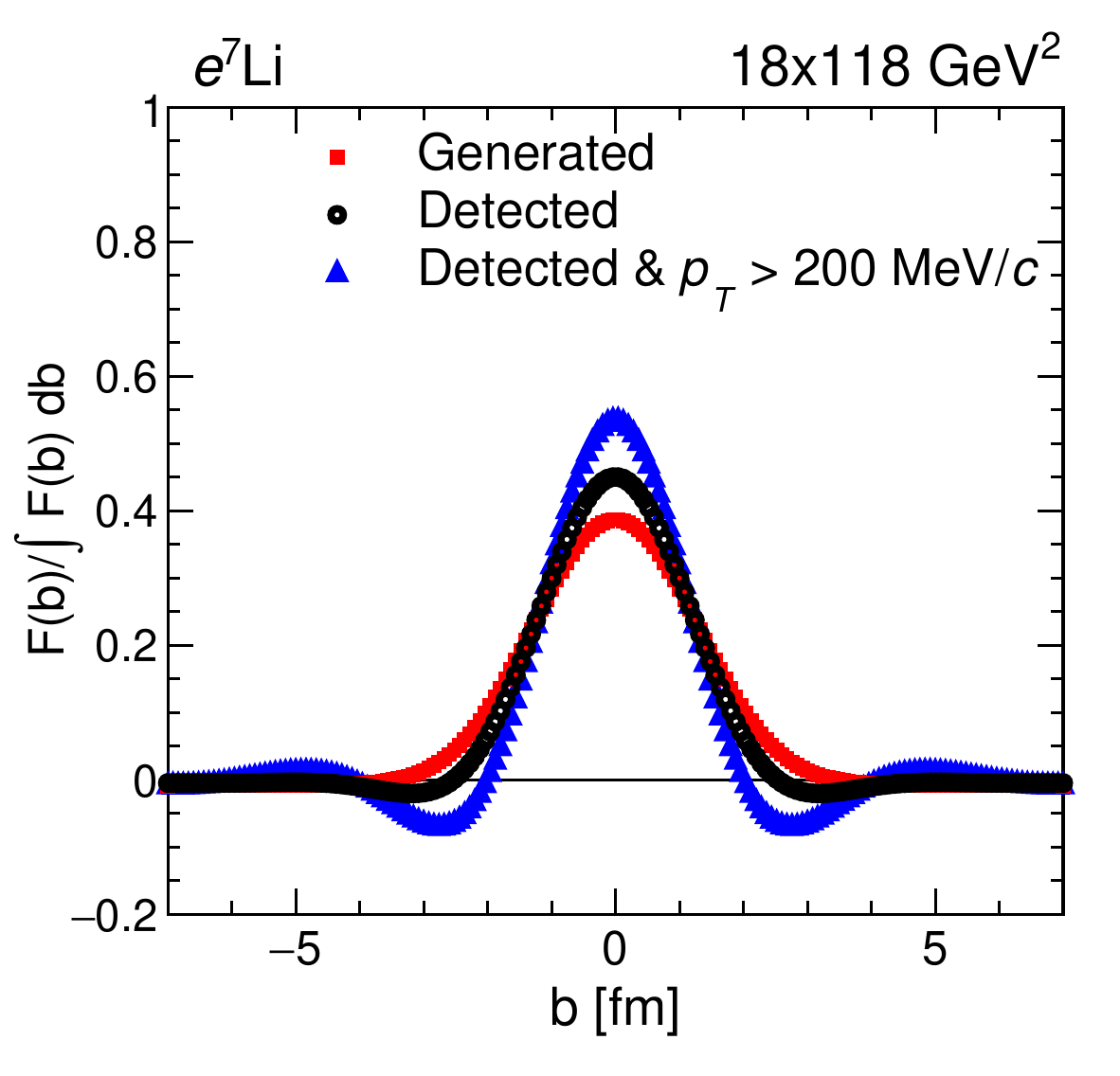}
\caption{The Fourier transforms from the distributions in Fig.~\ref{fig:t} for $e^{3}$He collisions (light panel) and $e^{7}$Li collisions (right panel) with top energies. Red squares represent the results at the MC-generated level. Black circles denote the Fourier transforms from the $|t|$ distributions of scattered light ions detected by the far-forward detectors at IR-8 (for $p_{T} >~0 ~\mathrm{MeV} /c$). Blue triangles stand for the Fourier transforms from the $|t|$ distributions of scattered light ions detected at IR-8 with a transverse momentum cut of $p_{T} >~200 ~\mathrm{MeV} /c$ (the underlying $|t|$ distribution is not shown in Fig.~\ref{fig:t}).}
\label{fig:b_Fb}
\end{figure*}

The Fourier transformation of the momentum transfer $|t|$ distribution of the coherent diffraction reveals the spatial distribution of transverse gluons. Therefore, it is critical for the proposed measurement to unambiguously identify the coherent process and measure its differential cross section as a function of $|t|$. 

Fig.~\ref{fig:t} illustrates the $|t|$ distributions for coherent $J/\psi$ production in $e^{3}$He collisions and $e^{7}$Li collisions at their respective top energies. Specifically, open red circles and open black squares represent the results of $e^{3}$He and $e^{7}$Li collisions in the MC-generated level, separately. In comparison, the filled markers indicate the distributions of the events detected by the far-forward detectors at IR-8 of EIC. At the bottom of the figure, the tagging efficiency distribution is shown as a function of $|t|$. The ratio of observable events shows an increase as $|t|$ increases. 

The coherent distributions in Fig.~\ref{fig:t} can be used to obtain information about the gluon distribution in impact parameter space $F(b)$ through a two-dimensional Fourier transform of the square root of the coherent elastic cross section~\cite{Toll:2012mb, Munier:2001nr}:
\be
F(b)=\frac{1}{2\pi}\int^{\infty}_{0}\mathrm{d} \Delta \cdot \Delta J_{0}(\Delta b)\sqrt{\frac{\mathrm{d} \sigma_{\rm coherent}}{\mathrm{d} |t|}}.
\ee
\noindent Here $\Delta=\sqrt{-t}$, $J_{0}$ is the Bessel function, and $\mathrm{d}\sigma_{\rm{coherent}}/\mathrm{d}|t|$ is the coherent differential cross section. In Fig.~\ref{fig:b_Fb}, we show the resulting Fourier transforms from distributions of both the MC-generated level and detected by the far-forward detectors at IR-8 in Fig.~\ref{fig:t}, using the range $\left | t \right | < 0.4 \mathrm{~GeV^{2}}$. The obtained distributions have been normalized to unity. As the detection efficiency for $e^{3}$He coherent processes is much better than that for $e^{7}$Li coherent processes, the distribution detected by the far-forward detectors is more consistent with that from the MC-generated level for $e^{3}$He collisions than for $e^{7}$Li collisions. 

In addition, we have studied a case where we select only scattered light ions with at least 200 MeV$/c$ of transverse momenta (not shown in Fig.~\ref{fig:t}). The corresponding impact parameter distributions from Fourier transforms of $^3\text{He}$ and $^7\text{Li}$ are shown in Fig.~\ref{fig:b_Fb}. The significant difference with respect to the full capability ($p_T\approx 0$ MeV$/c$) is to highlight the importance of low $p_T$ acceptance in the imaging program of light ions. The coherent ion tagging capability enabled by the optimized IR-8 design unlocks tangible achievements in broader nuclear exclusive processes. Specifically, it can be directly applied for Deeply Virtual Compton Scattering (DVCS)\cite{Ji:1996nm, Aschenauer:2025cdq} and Deep Exclusive Meson Production (DEMP)\cite{Tu:2023few, Horn:2017csb} on light nuclear targets, where precise identification of scattered coherent ions is critical for separating signal events from background. These processes are very sensitive to nuclear Generalized Parton Distributions (nGPDs)\cite{Aschenauer:2025cdq, Polyakov:2002yz} and nuclear transition GPDs\cite{Diehl:2024bmd}. These advantages increase the physics reach of IR-6 and strengthen EIC’s overall sensitivity to key nuclear observables.

\section{\label{sec:summary} Summary}

In this work, a study of the coherent light nuclei detection at the second interaction region at EIC is presented. eSTARlight is used to simulate the coherent vector meson production samples of $e+{\rm A}\rightarrow e'+{\rm A}'+\rm{VM}$ ($\rm{A= ~^{2}\rm{D}},~^{3}\rm{He},~^{4}\rm{He}, ~^{7}\rm{Li},~^{9}\rm{Be},~^{12}\rm{C}, ~^{16}\rm{O}$ and $\rm{VM}= \it{J/\psi}$, $\phi$, $\rho$) across different energies for both electron and nucleus beams. In the pre-conceptual design of the second interaction region, the incorporation of the secondary focus enhances the ability to tag particles with low transverse momentum and/or high longitudinal scattered ion momentum. This makes it possible to tag light ions with high rigidity, whose trajectory is largely collinear with the beam particles. 

Although a more comprehensive and realistic IR-8 design is necessary for future assessments of the advantages of incorporating a secondary focus, the present results indicate that enhancing the EIC exclusive, tagging, and diffractive physics program—specifically for light ions—is feasible through such an IR-8 design optimized for improved far-forward acceptance. In turn, this advancement will allow for cross-check of measurements obtained from IR-6 and will help achieve complementary fiducial acceptances among the various far-forward detectors across the two IRs.

\begin{acknowledgments}
The authors thank Jihee Kim, Bamunuvita Randika Gamage, Spencer Klein, and the local BNL group for valuable discussions. W. Chang is especially grateful to the BNL group for their warm hospitality during her visit, when this work was initiated. The work of W. Chang is supported by the National Natural Science Foundation of China with Grant No. 12305144. W. Chang is also supported by the Key Scientific Research Projects of Higher Education Institutions in Henan Province with Grant No. 24B140007. The work of E.C. Aschenauer, A.~Jentsch, and Z.~Tu is supported by the U.S. Department of Energy under Contract No.DE-SC001270, and A.~Kumar was supported by the Center for Frontiers in Nuclear Science. Z.~Tu is also supported by the BNL Laboratory Directed Research and Development (LDRD)23-050 project and 25-043 project. The work of Z. Yin is supported by the National Natural Science Foundation of China with Grant No. 12275103.
\end{acknowledgments}

\bibliography{bibliography}% Produces the bibliography via BibTeX.
\end{document}